\documentclass[twocolumn,aps,pre,showpacs,amsmath,amssymb,floatfix]{revtex4-2}
\usepackage{graphicx}
\usepackage{dcolumn}
\usepackage{bm}
\usepackage{xcolor}
\usepackage{soul}

\begin{document}

\title{Extracting Forces from Noisy Dynamics in Dusty Plasmas}

\author{Wentao Yu}
\email{wentao.yu@emory.edu}
\author{Jonathan Cho}
\author{Justin C. Burton}

\affiliation{Department of Physics, Emory University, Atlanta, Georgia 30322, USA}

\date{\today}
\begin{abstract}
Extracting environmental forces from noisy data is a common yet challenging task in complex physical systems. Machine learning (ML) represents a robust approach to this problem, yet is mostly tested on simulated data with known parameters. Here we use supervised ML to extract the electrostatic, dissipative, and stochastic forces acting on micron-sized charged particles levitated in an argon plasma (dusty plasma). By tracking the sub-pixel motion of particles in subsequent images, we successfully estimated these forces from their random motion. The experiments contained important sources of non-Gaussian noise, such as drift and pixel-locking, representing a data mismatch from methods used to analyze simulated data with purely Gaussian noise. Our model was trained on simulated particle trajectories that included all of these artifacts, and used more than 100 dynamical and statistical features, resulting in a prediction with 50\% better accuracy than conventional methods. Finally, in systems with two interacting particles, the model provided non-contact measurements of the particle charge and Debye length in the plasma environment.
\end{abstract}

\maketitle
\section{Introduction}
Huge amounts of experimental data are often collected faster than can be interpreted. In complex physical or biological systems, this data mostly comes in the form of tracked positions of individual agents or particles, yet random noise makes the inference of internal and external forces challenging. Conventional statistical methods often result in systematic error, and require special treatment of error estimation \cite{Lehle2015,Bruckner2020}. Machine learning (ML) algorithms can infer forces from trajectories without systematic error \cite{bongard2007automated,champion2019data,brunton2016discovering,daniels2015automated,Lusch2018,bapst2020unveiling,pathak2018model}, but their reported performance has been restricted to labeled simulated data rather than unlabeled experimental data. Another restriction is data mismatch \cite{vincent2017analysis,wang2021generalized}. Data is almost always simulated with Gaussian noise, which is presumed in many inference algorithms, but experimental data may include non-Gaussian noise and other artifacts such as systematic drift. Subsequently, modern inference algorithms should be benchmarked using experimental data where parameter estimates can be verified by alternative, independent methods. 


Dusty plasmas, where micron-sized charged particles are suspended in a low-density gaseous plasma, provide opportune experimental data for dynamical inference methods. The particles experience a wide array of forces including electrostatic repulsion, velocity-dependent drag from neutral and charged ions, and stochastic thermal kicks \cite{chaudhuri2011complex}. As a result, dusty plasmas display a wide range of complex, nonequilibrium dynamical phenomena, including superthermal excitations from non-reciprocal forces \cite{bockwoldt2014origin, ivlev2015statistical,qiao2015mode}, oscillations between ``turbulent'' and ``quiescent'' states \cite{gogia2017emergent,Gogia2020,kryuchkov2020strange}, parametric resonance and kinetic heating \cite{Williams2007,Norman2011,kong_qiao_matthews_hyde_2016}, spontaneous oscillations at low pressures \cite{Harper2020,Nunomura1999}, helical dust ``strings'' \cite{Kong2011,Hyde2013}, and vortical structure formation at high magnetic fields \cite{Choudhary2020,Schwabe2011,Thomas2016}. However, the individual interactions between particles are a subject of active research \cite{Sheridan2007,mukhopadhyay2012two,Lampe2000}, and the external forces acting on a single particle can be complex \cite{Melzer2019b,Harper2020,Nosenko2020,melzer2008fundamentals}. 

ML has already been applied to a few distinct areas of dusty plasma research. Examples include the interpretation of Langumir probe and electron density measurements \cite{ding2021probe,ding2021method,chalaturnyk2019first}, and the prediction of particle generation and annihilation in fusion devices \cite{bukhari2020design}. Additionally, ML has been used to identify phase boundaries in dense dust systems \cite{huang2019identification}, and to assist with stereoscopic tracking of many particles in three dimensions \cite{wang2020ML}. Bayesian analysis and ML have also been applied to investigate the nonlinear dynamics of single dust particles \cite{ding2021machine}. Importantly, these dynamics provide information about the dust charge, interaction potential, and external fields, essentially acting as a non-contact probe of the system. Both the dust charge ($Q$) and Debye screening length ($\lambda$) between interacting dust particles can by estimated by an analysis of the noisy dynamics \cite{mukhopadhyay2012two}, two parameters which are often difficult to accurately measure.

Here we show how the undisturbed, random motion of one and two particles in a dusty plasma can be interpreted using ML to provide accurate information about their inter-particle and environmental forces. Crucially, our ML methods are trained with simulations that consider real experimental artifacts such as anisotropic confinement, nonconservative forces, stochastic L\'{e}vy noise, non-Gaussian tracking error (pixel-locking), and experimental drift. These artifacts can be observed in a statistical analysis of the data, yet they rarely included in dynamical inference procedures, leading to data mismatch. In our procedure, features are extracted from the simulated trajectories to train supervised ML models. The models simultaneously predict system-wide parameters with 50\% better accuracy than traditional methods such as Fourier spectrum and maximum likelihood estimation in simulated trajectories. 

In the experiments, one key feature is that many of the parameters are independently inferred by analyzing the particles' recovery to equilibrium after a perturbation, thus labelling the data and verifying the model's performance on experimental time series. Based on labeling with this alternative method, our prediction on experimental data is evaluated to have the same accuracy as simulated data, alleviating data mismatch. Furthermore, in experiments with two particles, we provide an accurate estimation of $Q$ and $\lambda$ solely from the particles' pixel-scale Brownian motion without knowledge of other system-wide parameters, such as Epstein damping. These results will help guide other studies that use ML to quantitatively infer system parameters in real-world, noisy experimental data.

The rest of this paper will be organized as follows. In Sec.\ \ref{sec2}, we detail our experimental setup and 3D particle imaging and tracking methods. In Sec.\ \ref{1p_s}A, we introduce the linearized single particle model used for our simulations. Section \ref{1p_s}B explains the dominant source of errors in our experiments. Their mismatch from non-correlated Gaussian noise that is commonly used in simulations is observed by statistical analysis. We then explain how we handle these errors in our simulations. In Sec.\ \ref{1p_s}C, we describe the features extracted from simulated and experimental data. These features are used by our ML models. Section \ref{1p_s}D explains the different ML models and their corresponding performance on simulated test data compared with conventional methods. Section \ref{1p_s}E describes the alternative way that we label our experimental data. Finally, Sec.\ \ref{1p_s}F demonstrates the performance of our ML models on experimental, single particle data. Section \ref{2p_s} expands our methods to systems of 2 particles. In Sec.\ \ref{2p_s}A, we introduce the changes to the linearized model for 2 particles. Section \ref{2p_s}B and \ref{2p_s}C explains the simulation details and the features used for the ML models. Lastly, \ref{2p_s}D shows our predictions on experimental two particle data, including an inference of the particle charge and Debye length.

\section{Experimental Methods and Particle Tracking}

\label{sec2}

\begin{figure}[!]
\centering
\includegraphics[width=3.33 in]{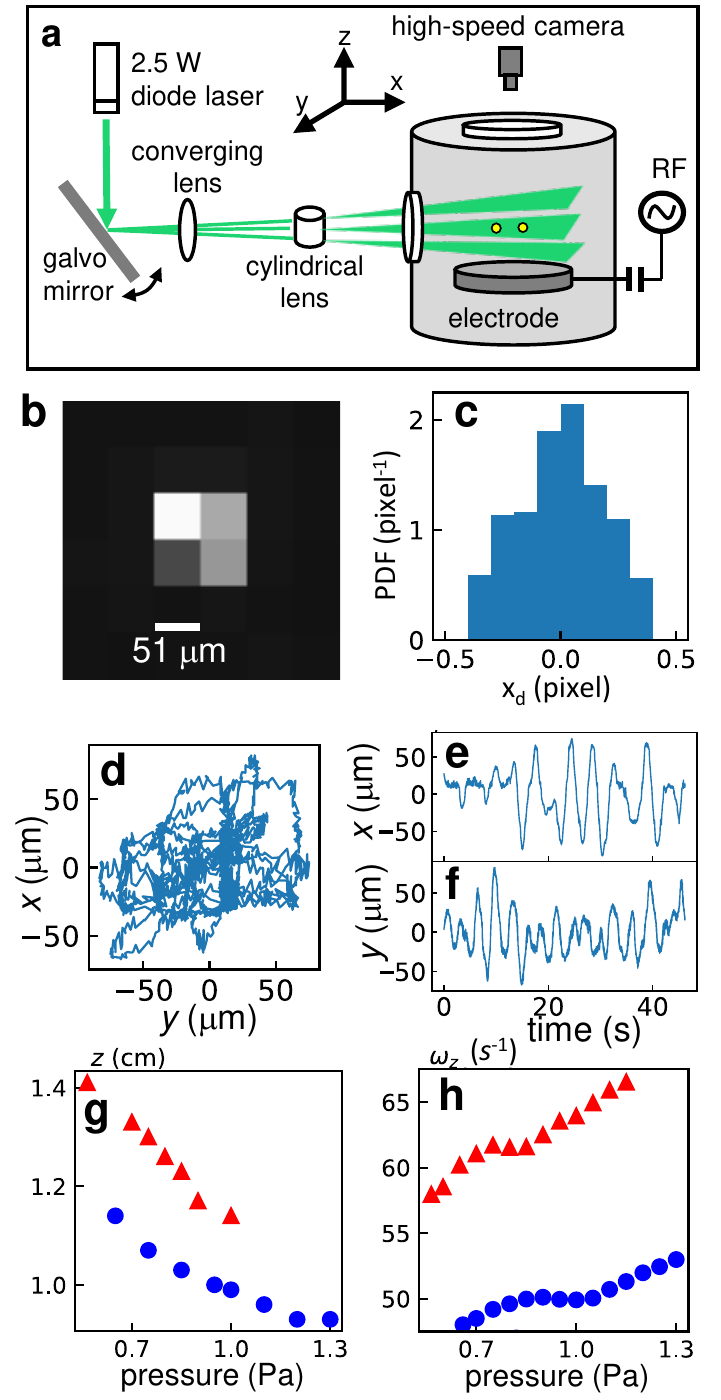}
\caption{(a) Experimental setup for the 3D tomographic imaging and particle tracking. The oscillating mirror varies the angle of the incoming laser (wavelength 532 nm). The converging lens focuses the beam in $z$, and the cylindrical lens expands the beam in the $xy$ plane. Particles are imaged and tracked from above as described in the text. (b) Image of the scattered light from a single particle with diameter 12.8 $\mu$m. (c) Probability distribution of the decimal part of tracked positions, prior to SPIFF correction. (d) A 45 s trajectory for a single 12.8 $\mu$m particle undergoing stochastic motion. (e-f) Time series of the $x$ and $y$ position corresponding to the same trajectory. (g) $z$-position as a function of pressure for 12.8 $\mu$m (blue circles) and 9.46 $\mu$m (red triangles) particles. (h) Dominant frequency of motion in the $z$-direction for both sizes of particles, obtained by Fourier transform.}
\label{p1_exp}
\end{figure}

Our experiments used melamine-formaldehyde (MF) particles with diameters 9.46 $\pm$ 0.10 $\mu$m and 12.8 $\pm$ 0.3 $\mu$m (microParticles GmbH). The particles were electrostatically levitated in a low-pressure argon plasma above an aluminium electrode with diameter 150 mm (Fig.\ \ref{p1_exp}a), similar to previous experiments \cite{gogia2017emergent,Harper2020}. The argon plasma was generated by a 13.56 MHz radio-frequency voltage applied to the electrode, resulting in 2.9 $\pm$ 0.1 W of input power and a fixed dc bias voltage of -36.3 $\pm$ 1.2 V. An aluminium ring was placed on the edge of the electrode to provide horizontal confinement. The plasma pressure, $P$, was varied between 0.6 Pa and 1.3 Pa. Under these conditions, the typical electron temperature in the plasma was 1.3-1.5 eV \cite{Harper2020}.  

To visualize the particles levitated in the plasma environment, a horizontal laser sheet was generated by focusing with a cylindrical lens, similar to previous experiments \cite{gogia2017emergent,Harper2020, Gogia2020}. The scattered light from the particles is captured from above by a Phantom V7.11 high-speed camera equipped with a macro lens. This allowed tracking the particle positions in the horizontal, $xy$-plane. Additionally, we used a mirror attached to a galvo motor to oscillate the laser sheet with a 50 Hz sawtooth wave at an amplitude of a few millimeters. The timebase of the camera was synchronized to the function generator driving the galvo, and the camera recorded at 1000 Hz (Fig.\ \ref{p1_exp}a), resulting in 20 images at different vertical positions per oscillation of the laser sheet. With this tomographic 3D tracking, we simultaneously obtained information about the vertical and horizontal motion of the particles. 

The spatial resolution of our imaging system was 51 $\mu$m per pixel in the $xy$-plane, and 200 $\mu$m between image slices in the $z$-direction. However, by tracking the 3D particle motion using an open source software (TrackPy \cite{allan_daniel_b_2021_4682814}), the position of the particles can be located with much better accuracy. The image representing the scattered light from a single particle is shown in Fig.\ \ref{p1_exp}b. The centroid of the particle ``blob" is found by calculating the center-of-mass of the pixels, where the pixel brightness represents the mass contribution of a single pixel \cite{allan_daniel_b_2021_4682814}. The same centroid procedure is done with image slices in the $z$-direction.  A probability density function of the decimal part of the tracked positions  ($x_d$) is shown in Fig.\ \ref{p1_exp}c, showing a strong bias towards integer values. This bias is known as pixel-locking \cite{feng2007}. Using the single-pixel interior filling function (SPIFF) algorithm \cite{burov2017single,yifat2017analysis}, these errors can be statistically corrected from the tracked data. Ultimately, our estimated sub-voxel resolution in tracking the particles was $\approx$ 4 $\mu$m in the $xy$-plane, and $\approx$ 16 $\mu$m in $z$. 

This sub-pixel error was confirmed using an independent procedure. We created digital movies of bright ``particles'' moving unidirectionally across a projection screen. The screen was imaged with our camera so that the particles appeared similar in size on the camera sensor when compared to the experiments (i.e. Fig.\ \ref{p1_exp}b). Since the trajectory of the particles was pre-determined in the movie, we compared the tracked positions to the known values. Despite these procedures, the horizontal resolution was still a significant fraction of the amplitude of the Brownian motion in the experiments. This was evidenced by systematic statistical effects in the analysis of the dynamics, and will be discussed in Sec.\ \ref{noise_sec}.

\section{Single Particle Motion}
\label{1p_s}
\subsection{The Linearized Model}
\label{1p_sA}

A typical $xy$ trajectory for a single, isolated particle is shown in Fig.\ \ref{p1_exp}d. The $x$ and $y$ time series corresponding to this trajectory is shown in Fig. \ref{p1_exp}e-f. Without any external perturbations, the particle experienced thermally-excited motion in three dimensions. The amplitude of the motion was $\approx$ 50 $\mu$m in the $xy$ plane. A dominant angular frequency of motion ($\omega\approx$ 1-2 Hz) is clearly visible in the time series. The amplitude of motion in $z$ was much smaller; less than our spatial resolution. Nevertheless, we measured the $z$-position of the particle as a function of gas pressure (Fig.\ \ref{p1_exp}g), which increased at lower pressure as the electrode's sheath expanded. Also, by Fourier transforming the time series of the $z$-position, we estimated the vertical frequency of oscillation ($\omega_z$, Fig.\ \ref{p1_exp}h), which was much larger than the horizontal frequency, indicating strong confinement in the $z$-direction. 

\begin{figure}[!]
\centering
\includegraphics[width=3.33 in]{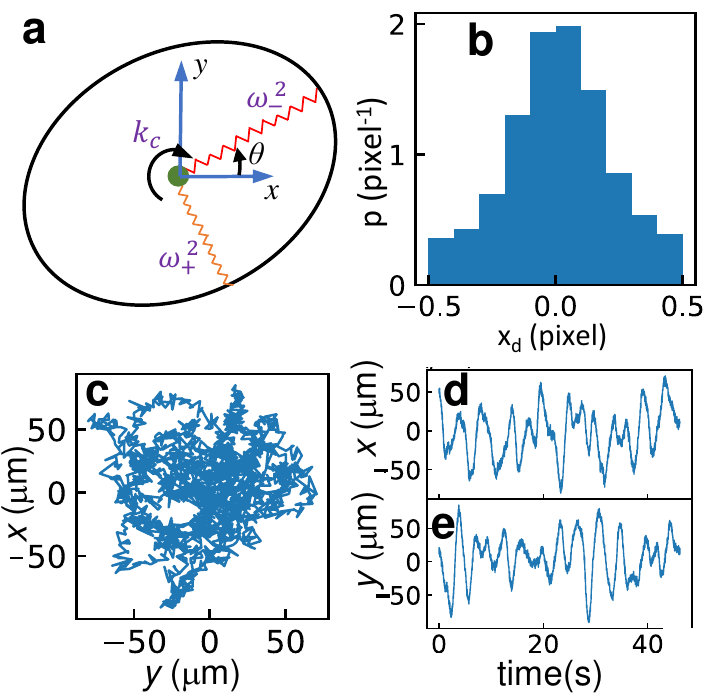}
\caption{(a) Linearized external forces on the particle, as described in the text. The black circle represents an equipotential surface of the harmonic confinement. The eccentricity is exaggerated. (b) Probability distribution of the decimal part of positions after simulated pixel-locking. (c) A simulated trajectory of length 45 s for a single particle undergoing stochastic motion. (d-e) Time series of the $x$ and $y$ position corresponding to the trajectory shown in (b). }
\label{p1_sim}
\end{figure}

Due to the small amplitude of motion, to lowest order, the particles behaved as stochastic harmonic oscillators. Since the amplitude of motion in $z$ was much smaller due to the strong confinement, we will ignore motion in the $z$-direction for our linearized model. Let $\vec{\mathbf r}(t) = x(t) \vec{\mathbf e}_x+ y(t)\vec{\mathbf e}_y$ denote the two-dimensional (2D) position of a particle at time $t$, and dotted variables refer to time derivatives. The linearized dynamics of one particle reads: 
\begin{equation}
   \Ddot{\vec{\mathbf r}}  = -\vec{\mathbf \nabla}\phi + \vec{\mathbf \nabla} \times \vec{\mathbf A} -\gamma\dot{\vec{\mathbf r}} + \vec{\mathbf N}(\alpha)
    \label{eq1}
\end{equation}
\begin{equation}
\begin{aligned}
    \phi = &\frac{\omega^2}{2}[(1-\delta)(x\cos\theta + y\sin\theta)^2 + \\&(1+\delta)(-x\sin\theta + y\cos\theta)^2]
\end{aligned}
\label{eq2}
\end{equation}
\begin{equation}
   \vec{\mathbf A} = \frac{k_c}{2}(x^2+y^2)\vec{\mathbf e}_z 
  \label{eq3}
\end{equation}
This model contains 6 parameters, $\gamma$, $\omega$, $\delta$, $\theta$, $k_c$, and $\alpha$. The deterministic confinement force has two components. The conservative potential $\phi$ resembles a 2D spring characterized by 3 parameters: the average frequency $\omega$, the asymmetry $\delta$ between two principal axes, and the angle $\theta$ from the $x$-axis to the weaker principle axis. Two eigenfrequencies, $\omega_- = \omega \sqrt{(1-\delta)}$ and $\omega_+ = \omega \sqrt{(1+\delta)}$, and $\theta$ are displayed in Fig.\ \ref{p1_sim}a. The nonconservative vector potential $\vec{\mathbf A}$, characterized by a parameter $k_c$, represents a rotational force possibly due to drag from the background ion flow \cite{Chai2016}, particle asymmetries \cite{Nosenko2020}, or magnetic fields \cite{Konopka2000}.

Additionally, the particle experiences drag from the background neutral gas. According to Epstein's law assuming diffuse reflection from neutral gas collisions on the particle surface, the damping coefficient can be expressed as \cite{melzer2008fundamentals,Harper2020}:
\begin{equation}
   \gamma = 1.44 \frac{P}{a_p\rho_p}\sqrt{\frac{2m_{ar}}{\pi k_BT}}.
    \label{epstein}
\end{equation}
Here, $P$ is the gas pressure, $a_p$ is the particle radius, $\rho_p = 1510$ kg/m$^{3}$ is the mass density of the particle, $m_{ar}$ is the mass of an argon atom, $k_B$ is Boltzmann constant, and $T = 300$ K is room temperature. An important assumption here is that the particle size is smaller than the mean free path ($\approx$ 5 mm at $P$ = 1 Pa). For the particles with diameter $2a_p = 12.8$ $\mu$m, $\gamma/P$ = $0.95$ Pa$^{-1}$ s$^{-1}$. For the particles with diameter $2a_p = 9.46$ $\mu$m, $\gamma/P$ = $1.29$ Pa$^{-1}$ s$^{-1}$.


\subsection{Handling Random Noise, Parameter Drift, and Measurement Error in simulation}
\label{noise_sec}

The last term in Eq.\ \ref{eq1} is a stochastic acceleration, $N(\alpha)$, which follows a stable L\'{e}vy distribution. The parameter $\alpha$ will be determined by inference. We do not assume a priori that the stochastic motion is purely Brownian ($\alpha = 2$), and $\alpha<2$ indicates a more heavy-tailed distribution. The Brownian motion of particles in experiments is driven primarily by random impulses from the environment. Thus, in the simulations, temporally-independent random noise is added to the $acceleration$ of the particles at each time step. The L\'{e}vy-stable noise was generated by the python function \textit{scipy.stats.levy\_stable} with parameter skewness $\beta$ = 0 and center $\mu$ = 0. The noise scale $c$ and the parameter for heavy-tailness $\alpha \in (1.6,2.0)$ was independently chosen for each simulation. 

For all simulations, we used a time step $\Delta t$ = 0.02 s to closely follow the experiments. The parameters $\omega$, $\gamma$, $\delta$, $\theta$, $k_c$, and $\alpha$ are randomly chosen from a uniform distribution prior to each simulation. The range of values possible for each parameter are listed in Table \ref{tab1}. For the maximum values of $k_c$, we chose $k_{c,max}$ = $\min(1 \text{ s}^{-2},0.9\times \sqrt{\omega^4\delta^2 + \omega^2\gamma^2})$, which guaranteed that the confinement force was able to provide the necessary centripetal acceleration to keep the particle in a bounded stable orbit. Since none of the parameters have a length scale in their units, the simulated L\'{e}vy noise scale was arbitrarily set to $c = \sqrt{\gamma\omega^2/\Delta t}$. 

\begin{figure}[!]
\centering
\includegraphics[width=\columnwidth]{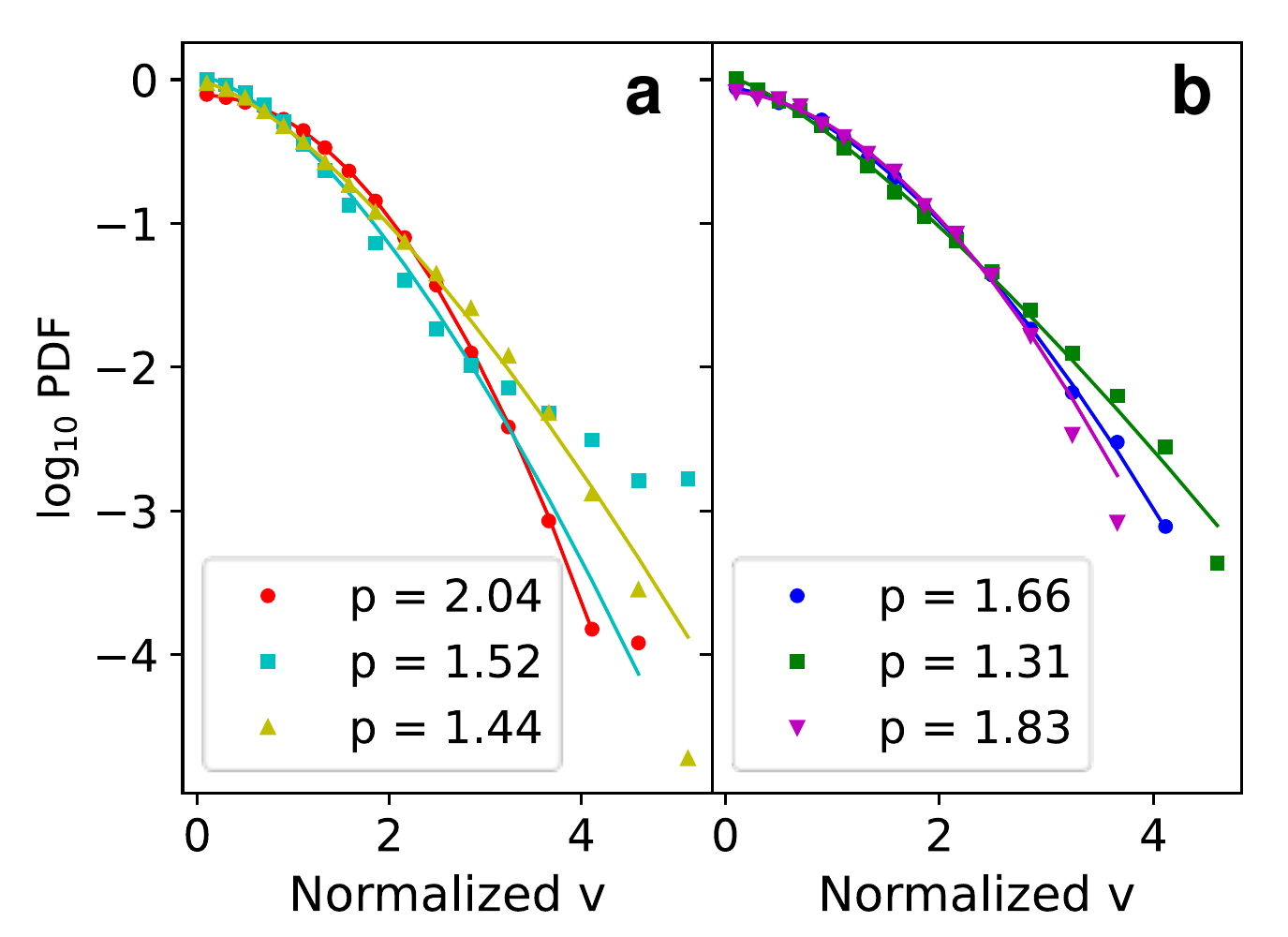}
\caption{Probability density function (PDF) of the $x$-component of the velocity, $|v_x|$, normalized by $\sqrt{\left<v_x^2\right>}$, where the average is over time. (a) The velocity distribution of simulated data. Red circles represent $\alpha = 2$ (Gaussian noise). Green squares represent $\alpha=1.8$ (Non-Gaussian noise). Yellow triangles represent $\alpha = 2$, but with simulated pixel-locking and SPIFF correction. The solid lines are fits to the form $y = Av^p$ with more weight attached to the left side of the curve (see Sec.\ \ref{features_single}, part 4). The fitted value of the exponent $p$ is shown in the inset. (b) 3 different velocity distributions for experimental trajectories and the associated fits with exponent $p$. All 6 trajectories in the curve undergoes a same low-pass filter with a  4 Hz cutoff.}
\label{vdis}
\end{figure}

Drift was inevitably present in nearly all experiments. This was most noticeable in the drift of the equilibrium position of the particle. The drift was small, less than 1 pixel, but is still comparable to the amplitude of the Brownian motion. We modeled this in simulations as a temporally-correlated Gaussian noise added to the $equilibrium$ $positions$. To simulate a time series of temporally-correlated noise, ${a_i}$, with standard deviation (STD) $\sigma$ and characteristic correlation time $\tau$ much larger than simulation time step, $\tau>>\Delta t = 0.02$ s, we used the recursive relation:
\begin{equation}
    \begin{aligned}
    a_0 &= N_0\\
    a_i &= \left(1-\frac{\Delta t}{\tau}\right)a_{i-1} + \frac{\Delta t}{\tau}N_i.
    \end{aligned}
    \label{drift_eq}
\end{equation}
Here $N_i$ is an array of independent and identically-distributed Gaussian random numbers with zero mean and unit variance. The final sequence is adjusted by subtracting the mean from each element in the series, and then normalizing the STD to be $\sigma$. During simulation, the equilibrium positions ($x_0$ and $y_0$) both drift with the same timescale  $\tau \in (12,800)$ s and potentially different amplitudes $\sigma \in (0,0.5)$, randomly chosen for each simulation (Eq.\ \ref{drift_eq}). 

As discussed in Sec.\ \ref{sec2}, pixel-locking was an important source of noise in experimental data. Thus, measurement errors were added to the simulated position time series after all the time steps were $completed$. This was intended to simulate errors associated with tracking the particles in the images. 
To simulate pixel-locking in the position time series, we converted the simulated position to pixels using a length scale $l_{pixel}$ and a random offset $x_{pixel} \in (-0.5,0.5)$. Then we applied a transformation to the decimal portion, $x_d\in (-0.5,0.5)$, of the pixel:
\begin{equation}
    x_d = \frac{x + x_{pixel}}{l_{pixel}} - \text{round}\left(\frac{x + x_{pixel}}{l_{pixel}}\right),
\end{equation}
\begin{equation}
    x_d^* = sgn(x_d) \times \frac{|2x_d|^{p_d}}{2} + N_t(\sigma_d).
\end{equation}
Here $x_d^*$ is the transformed pixel value, $p_d$ is an exponent randomly chosen between $(1,4)$ for each time series, and $N_t$ is a Gaussian noise with zero mean and STD $\sigma_d \in (0,0.1)$. The distribution of the decimal part of simulated `pixels' $x_d$ is plotted in Fig.\ \ref{p1_sim}b. Finally, as in the experiments, we used the single-pixel interior filling function (SPIFF) algorithm \cite{yifat2017analysis} on $x_d^*$ to correct simulated  data before training the model. An example of a simulated trajectory and its $x$ and $y$ components are shown in Fig.\ \ref{p1_sim}c and \ref{p1_sim}d-e, respectively.

Though pixel-locking was a small source of error, it led to large systematic errors in dynamical quantities such as the 1D velocity distribution. Without further modeling the effects of pixel-locking, these errors can be easily mistaken for stochastic noise with $\alpha<2$. In a stochastic under-damped harmonic oscillator simulated with $\alpha = 2$ and Gaussian measurement error, the 1D velocity distribution was well-fit by the form $\log P(v_x)=Av_x^p$ with $p = 2$ (Fig.\ \ref{vdis}a, red circles). However, a simulation with either a smaller value of $\alpha$ (green squares), or pixel-locking (yellow triangles) both led to a significantly smaller value of the fitted parameter $p$. Experimentally, the fitted $p$ was usually smaller than 2 (Fig.\ \ref{vdis}b). It is possible to minimize pixel-locking errors in the velocity distribution function by defocusing the camera \cite{feng2007,mukhopadhyay2012two}, however, our 3D imaging and tracking methodology required particles with significant brightness due to the low exposure time. Subsequently, it was not possible to determine whether $\alpha < 2$ or pixel-locking leads to non-Gaussian distributions with our current analysis.

\begin{table*}
\caption{The parameters for 1-particle simulation.}
\label{tab1}
\begin{center}
\begin{tabular}{|c| c| c |c | c|} 
 \hline
 Name & Description & Range & Drift amp. & Drift time \\ 
 \hline\hline
 $\omega$ & confinement freq. & (1.3,2.5) s$^{-1}$ & None & None \\ 
 \hline
 $\gamma$ & damping coef. & (0.4,1.7) s$^{-1}$ & None & None \\
 \hline
 $\delta$ & asymmetry & (0,0.35) & None & None \\
 \hline
 $\theta$ & weak axis & $(-\pi/2,\pi/2)$ & None & None \\
 \hline
 $k_c$ & vortex force coef. & $(-k_{c,max},k_{c,max})$ s$^{-2}$ & None & None \\ 
 \hline
 $\alpha$ & noise distribution & (1.6,2.0) & None & None\\
 \hline
 $x_0$ & equi. position & 0 & (0,0.5) & (12,800) s\\
 \hline
 $y_0$ & equi. position & 0 & (0,0.5) & (12,800) s\\
 \hline
 $l_{pixel}$ & simulated pixel width & (0.3,1) & None & None\\
 \hline
\end{tabular}
\end{center}
\end{table*}



\subsection{Features for ML}

\label{features_single}

The data used to train the ML model consisted of simulated time series of both the $x$ and $y$ motion of the particle. Typically, each time series contained 15,000 to 100,000 elements, depending on the total length of time of the motion. Although in principle it is possible to use the raw data as input to the ML model, this would drastically increase the computation time. Thus, we choose to extract hundreds of relevant dynamical features of the motion in order to train the model. These ranged from Fourier transforms and autocorrelations, to more sophisticated inference algorithms such as underdamped Langevin inference (ULI) \cite{Bruckner2020}.

The confining potential for the particles consists of an asymmetric harmonic trap in $x$ and $y$, as shown in Fig.\ \ref{p1_sim}a. We first extracted a rough estimation of the principle axes, defined by $\theta$, for a 2D time series $[x_t, y_t]$, $t = 0,1,2,\ldots,T$. $T$ is the length of a single time series, and is the first feature. The total time duration of the series is $T\times\Delta t$, where $\Delta t$ = 0.02s. In polar coordinates, $\phi_t = \arctan \frac{y_t}{x_t}$. We used 20 bins to form a histogram of $\phi_t$ between $(-\pi/2, \pi/2)$ and fit the probability density with
\begin{equation}
p(\phi) = \frac{1 + \delta_{hist} \cos2(\phi -\theta_{hist})}{\pi}.
\end{equation}
Here $\delta_{hist}$ and $\theta_{hist}$ are two features.

Let $\left<p_i,q_i\right> = \Sigma_i^Tp_iq_i/T$. The correlation matrix $\textbf{C}$ was computed: 
\begin{equation}
    \textbf{C} = 
    \begin{bmatrix}
        \left<x,x\right> & \left<x,y\right>\\
        \left<x,y\right> & \left<y,y\right>
    \end{bmatrix}.
\end{equation}
The eigenvalues of the matrix are $\lambda(1-\delta_{eig})$ and $\lambda(1+\delta_{eig}$ and their eigenvectors are ($\cos\theta_{eig}, \sin\theta_{eig}$) and ($-\sin\theta_{eig},\cos\theta_{eig}$). Here $\lambda$, $\delta_{eig}$, and $\theta_{eig}$ are three features. 
After calculating the eigenvectors, ($x_t,y_t$) are projected onto the (estimated) weaker and stronger principle axes for further analysis:
\begin{equation}
    \begin{aligned}
    w_t = x_t\cos\theta_{eig} - y_t\sin\theta_{eig}\\
    s_t = x_t\sin\theta_{eig} + y_t\cos\theta_{eig}
    \end{aligned}
\end{equation}
$w_t$ and $s_t$ are then normalized into unit STD, and the following feature extraction algorithms are applied to $(w_t,s_t)$:
\begin{enumerate}
    \item Fourier spectrum. This is the most commonly used tool to analyze the motion of a 1D harmonic oscillator, $\xi_t$. We compute the Fourier spectrum and only analyzed data between 0.5 s$^{-1} < \omega < $ 4 s$^{-1}$. This is fitted to analytical prediction for a 1D stochastic harmonic oscillator: 
    \begin{equation}
        \omega I(\omega) = A_{FT}\left[\omega^2\left(1-\frac{\omega_{FT}^2}{\omega^2}\right)^2 + \gamma_{FT}^2\right]^{-1/2}
    \label{FT}
    \end{equation}
    where $A$, $\omega_{FT}$ and $\gamma_{FT}$ are fitting parameters. Although $\theta_{eig}$ is a good estimate of the principal axes, we performed Fourier analysis on a combination of $w_t$ and $s_t$: $\xi_t = w_t \cos \phi+s_t\sin\phi$. The following pseudocode describes the procedure:\\\\
   for $\phi$ = [$-\pi/4$, 0, $\pi/4$, $\pi/2$]:\\
       \hphantom{~~~~} $\xi_t = w_t\cos\phi + s_t\sin\phi$,\\
       \hphantom{~~~~} Conduct 1D Fourier spectrum on $\xi_t$,\\
       \hphantom{~~~~} Fit the spectrum using Eq.\ \ref{FT},\\
       \hphantom{~~~~} $A_{FT,\phi}$, $\omega_{FT,\phi}$ and $\gamma_{FT,\phi}$ are features.\\\\
    Altogether 12 features are extracted using the Fourier spectrum. 
    
    \item Autocorrelation is another analysis technique used on 1D time series, $\xi_{t}$, and is defined as:
    \begin{equation}
        A(\tau) = \sum_{t=0}^{T-\tau}\xi_t\xi_{t+\tau}/(T-\tau).
        \label{A0}
    \end{equation}
    $A(\tau)$ was computed for $\xi_t$ and fitted to the analytic form for a 1D stochastic harmonic oscillator:
    \begin{equation}
        A(\tau) = \left(1+\frac{\gamma_{A}^2}{\omega_{A}^2}\right)e^{-\gamma_{A}\tau}\cos\left(\omega_{A}\tau - \arctan\frac{\gamma_{A}}{\omega_{A}}\right)
        \label{A1}
    \end{equation}
    Similar as Fourier spectrum, the following loop is performed to extract features.\\\\
        for $\phi$ in [$-\pi/4$, 0, $\pi/4$, $\pi/2$]:\\
        \hphantom{~~~~} $\xi = w\cos\phi + s\sin\phi$,\\
        \hphantom{~~~~} Normalize $\xi$ into zero mean and unit variance
        \hphantom{~~~~} Calculate the autocorrelation by Eq.\ \ref{A0},\\
        \hphantom{~~~~} Fit the autocorrelation using Eq.\ \ref{A1},\\
        \hphantom{~~~~} $\omega_{A,\phi}$ and and $\gamma_{A,\phi}$ are features.\\\\
    Altogether 8 features are extracted using autocorrelation.
    \item Percentiles and equipartition law. 
    Let $P(\xi,p)$ indicate the $p$ percentile of a 1D time series $\xi_t$, the quantity $\zeta = \frac{P(\xi,1) - P(\xi,99)}{P(\xi,30) - P(\xi,70)}$ contains qualitative information about the heavy-tailness of the distribution of the stochastic noise that drives $\xi$. Furthermore, according to equipartition, the time-averaged kinetic and potential energies should be equal. As a result, 
    \begin{equation}
        \omega_{ep}^2 = \frac{\sum_{t=1}^{T-1}\xi_t^{\prime 2}}{\sum_{t=1}^{T-1}\xi_t^2}
    \end{equation} 
    is a rough estimation of the eigenfrequency if ${\vec{\mathbf e}}_\xi$ is a principle axis for the confinement, where 
    $\xi^\prime_{t} = \frac{\xi_{t+1} - \xi_{t-1}}{2\Delta t}$ and $\Delta t = 0.02s$.
    To extract features, the following loop is performed.\\\\
        for $\phi$ in [$-2\pi/3$, $-\pi/3$, 0, $\pi/3$, $2\pi/3$, $\pi/2$]:\\
        \hphantom{~~~~} $\xi = w\cos\phi + s\sin\phi$,\\
        \hphantom{~~~~} Calculate $\zeta_\phi$ and $\omega_{ep,\phi}$ as features.\\\\
    Altogether 12 features are extracted. 
    
    \item Velocity distribution. For a 1D time series $\xi_t$, the central difference velocity is calculated, $\xi^\prime_t$. Then we compute the probability distribution $P(\xi^\prime_n)$ of $\xi^\prime_n=|\xi^\prime_t|/\sqrt{\left<\xi^{\prime 2}_t\right>}$, as done in Fig.\ \ref{vdis}. Were the noise purely Gaussian ($\alpha = 2$) with no measurement error, then $\log(P(\xi^\prime_n))\propto -\xi^{\prime 2}_n$. A more heavy-tailed distribution (see Fig.\ \ref{vdis}) may indicate $\alpha < 2$ or pixel-locking measurement error. Since the distribution is rather complicated, three different fits are performed to extract features. The first is a fit of $\log P$ versus $\xi^\prime_n$:
    \begin{equation}
        \log P(\xi^\prime_n) = A_0(\xi^\prime_n)^{p_0},
    \end{equation}
    where a fitting weight, $e^{\frac{\log P}{2}}$, is applied to attach more importance to the beginning of the curve. $A_0$ and $p_0$ are fitting parameters. The second fit linearly fits the $\log P$ vs. $\xi^\prime_n$ curve with $\xi^\prime_n>2.5$. The linear coefficient $p_1$ is recorded. 
    The third fit linearly fits $P$ VS $\xi^\prime_n$ with $\xi^\prime_n > 2.5$. The linear coefficient $p_2$ is recorded. Note that the second and third fits lack physical meaning, but they provide some qualitative information that helps the ML model give quantitative predictions. The following loop was used:\\\\
    for $\phi$ in [$-\pi/12$, $\pi/4$, $7\pi/12$]:\\
        \hphantom{~~~~} $\xi = w\cos\phi + s\sin\phi$,\\
        \hphantom{~~~~} Plot the histogram of $P(\xi^\prime_n)$\\
        \hphantom{~~~~} Fit histograms to get features $p_{0,\phi}$, $p_{1,\phi}$, $p_{2,\phi}$.\\\\
    Altogether 9 features are extracted.

    \item Intermittency analysis. For a 2D time series $(w_t,s_t)$, a scalar velocity  is defined as 
    \begin{equation}
        \xi^\prime_{t} = \frac{\sqrt{(w_{t+1} - w_{t-1})^2 + (s_{t+1} - s_{t-1})^2}}{2\Delta t}
    \end{equation}
    and its average over time $\tau$ is defined as
    \begin{equation}
        \bar{\xi}^\prime_{i}(\tau) = \sum_{t = i\tau+1}^{(i+1)\tau}\frac{\xi^\prime_{t}}{\tau} 
    \end{equation}
    where $i$ = $0,1,2,\ldots,\lfloor \frac{T-1}{\tau}-1\rfloor$.
   We introduced this particular measure because at relatively high values of the vortical force amplitude, $k_c$, the particle displays intermittent behavior characterized by large orbital excursions from equilibrium, yet below the critical value of $k_{c,max}$. As a result, the $\xi^\prime$ fluctuates at lower frequencies than other time scales in this system. This is characterized by the standard deviation (STD) of $\bar{\xi}^\prime_{i}(\tau)$ over $i$. Two features are extracted from the 2D trajectory with $\tau$ = 500 and 1300, respectively. 
    
    \item Vorticity estimation. Given a 2D time series $(w_t,s_t)$, the 2D velocity vector $\vec{\bf v}_t$ is first calculated. Let $v_{mean} = \sqrt{\left<v^2_t\right>}$ where $\left<\cdots\right>$ represents averaging over $t$. A qualitative estimation of angular velocity is used:
    \begin{equation}
        \Omega(\tau) =\frac{1}{\Delta t} \left<\frac{\vec{\textbf v}_t\times \vec{\textbf v}_{t+\tau}}{v_{mean}\left(\frac{v_{t} + v_{t+\tau}}{2}\right)\tau}\right>.
    \end{equation}
    This form puts a larger weight on larger velocities, which is necessary since $\Omega$ is completely dominated by noise for small velocities. $\Omega(1)$ and $\Omega(5)$ are two features used in the model.
    
    \item Linear correlation and mutual information. Built-in python packages \cite{scikit-learn} are used to compute the linear correlation and mutual information between all pairs of time series ($w$, $s$), ($w$, $v_w$), ($w$, $v_s$), ($s$, $v_w$), ($s$, $v_s$), where $v_w$ ($v_s$) is the central difference velocity associated with $w$ ($s$). These measurements are most relevant for large values of $k_c$, where circular motion can be detected. Altogether this provides 10 features. 
    
    \item Underdamped Langevin Inference (ULI) \cite{Bruckner2020}. ULI is a maximum-likelihood algorithm based on modified linear-regression. The time series $w$, $s$, $v_w$, and $v_s$ are used as inputs, along with a linear model of the forces, and the parameters of interest (i.e. $\omega$, $\gamma$, etc.) are estimated. There are 8 coefficients in the linear regression, which are 8 features. 
    
    \item The previous 8 analyses gives 63 features. Then, a band-pass filter is applied to $w_t$ and $s_t$ with an upper threshold = 2 Hz and a lower threshold = 0.01 Hz. The previous 8 analyses are repeated for 63 more features. This was done to reduce noise in the original data, yet by including analysis on both filtered and unfiltered data, we avoid losing information with little cost of adding features. Counting the 6 features in preprocessing, altogether there are 132 features for the motion of single particles.
    \end{enumerate}
    
\subsection{ML Methods and Performance}

\begin{figure}[!]
\centering
\includegraphics[width =3.2in]{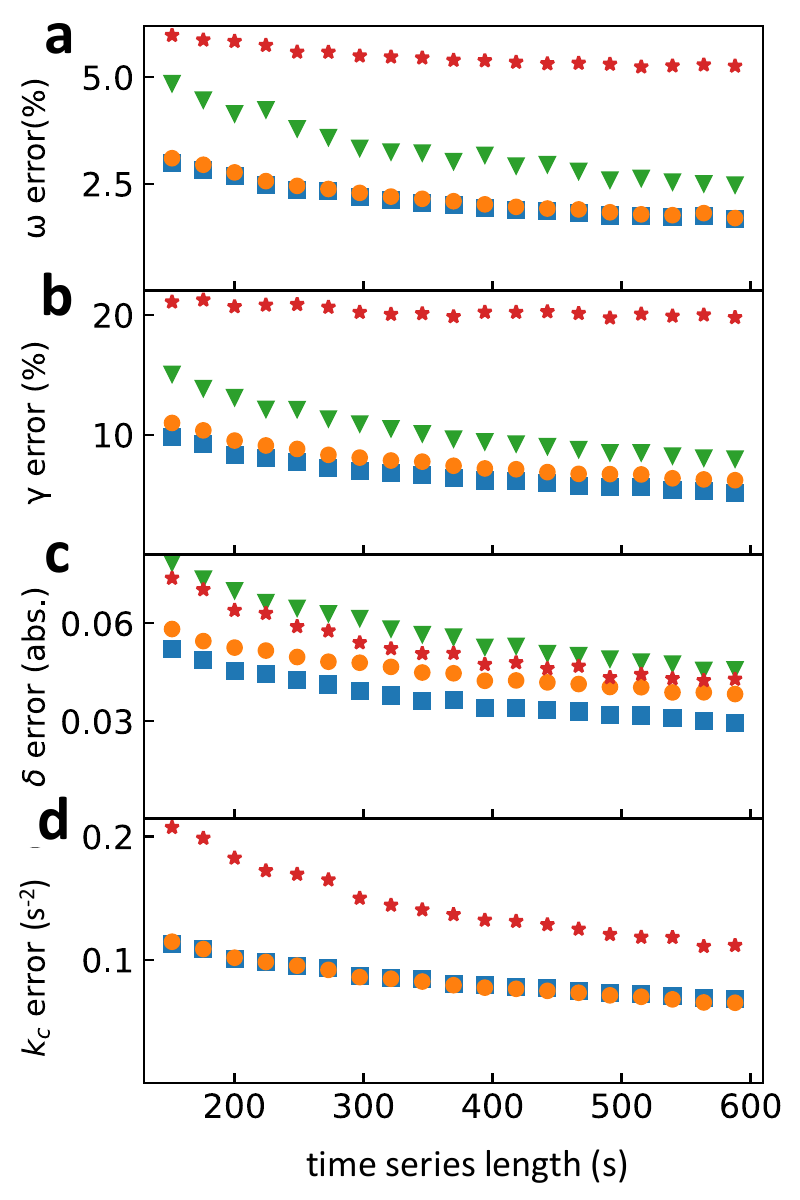}
\caption{The prediction error of various models for multiple parameters on one particle simulated test data. Red stars represent ULI, green triangles represent Fourier spectrum, blue squares represent neural network, and orange circles represent gradient boosting. Note that the Fourier spectrum cannot predict $k_c$ and must be based on a known $\theta$. Since ML is trained on a certain range of all parameters (Table \ref{tab1}), unreasonable predictions of Fourier spectrum and ULI are also cropped to that range.}
\label{supp3}
\end{figure}

\begin{figure*}[!]
\centering
\includegraphics[width =\textwidth]{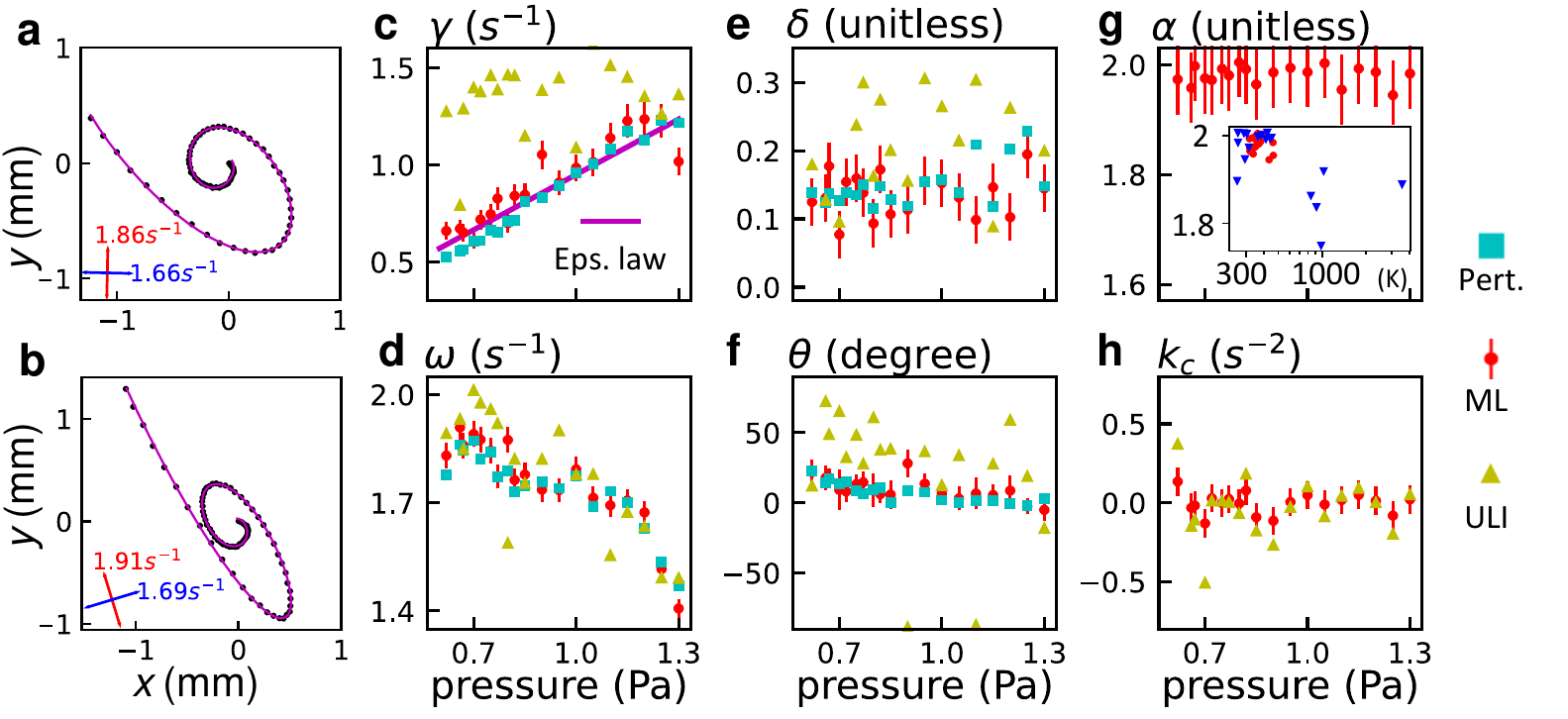}
\caption{(a-b) Two different experiments of the same particle relaxing to equilibrium after a perturbation. The pressure was $P$ = 0.80 Pa. The magenta lines are fits using Eqs.\ \ref{perturbfit3}-\ref{perturbfit4}. The red and blue lines indicate $\omega_+$ and $\omega_-$ and their orientations, respectively. 
(c-f) The prediction from ML (the mean of the predictions from neural network model and gradient boosting model, red circles), ULI (yellow triangles), and reference estimation from the perturbation experiments (Pert., cyan squares) for $\gamma$, $\omega$, $\delta$, and $\theta$ for particles with diameter 12.8$\mu$m. The purple line in (c) represents the theoretical value of Epstein's Law (Eq.\ \ref{epstein}). (g) $\alpha$ and (h) $k_c$  as predicted by ML. These parameters cannot be verified by the perturbation experiments. The inset in (g) shows the prediction of $\alpha$ correlates with the potential temperature of the particle. Red squares represents 12.8 $\mu$m particles and blue triangles represent 9.46 $\mu$m particles. Error bars were obtained from predictions on the simulated test data set, and errors based on fitting perturbed trajectories are smaller and not shown for clarity.} 
\label{p2}
\end{figure*}

\textcolor{black}{Two python-based ML algorithms (gradient boosting, an ensemble of decision trees, and neural network \cite{scikit-learn}) were trained on 132 extracted features from 400,000 simulated time series (training data set) to predict the 6 randomly-chosen parameters. Within the algorithms, the gradient boosting model has parameters n\_estimators = 250, max\_depth = 5, and the dense neural network has 5 hidden layers, with size (128,64,64,32,16) and all hyperbolic tangent activations. Before training, both the features and the targets are normalized by the whole training batch to zero mean and unit variance. The performance of each method was benchmarked on 80,000 simulated time series (test data set). Figure \ref{supp3}f shows that ML methods are $\approx 1.5\times$ more accurate at predicting $\omega$ and $\gamma$ than simply fitting analytical expressions to the Fourier spectrum of the data along the principal axes of confinement, and 2-3$\times$ more accurate than ULI \cite{Bruckner2020}.} 

\textcolor{black}{With regard to Underdamped Langevin Inference (ULI), we note that the performance was excellent and comparable to the prediction error for ML \textit{when using only Gaussian noise, no pixel-locking, and no drift}. These sources of noise seemed to dramatically reduce the performance of ULI, yet these sources of noise are unavoidable in real experimental data. However, despite it's lack of parameter estimation power on single, noisy data sets, ULI consistently ranked as one of the most important predictive features in the ML algorithms. Employing ULI in the simulated features increased the total simulation and feature extraction time by 150\%.}

\subsection{Labeling Experimental Data}

It is challenging to verify the accuracy of results when applying ML models to unlabeled experimental data. However, in our experiments, we measured the parameters using an independent, alternative method. By perturbing the particle with a magnet outside the chamber and observing the particle's relaxation to equilibrium, we fit the 2D trajectory and obtained estimates of $\omega$, $\gamma$, $\delta$, and $\theta$. \textcolor{black}{Initially, we used a ``mechanical" method to peturb the particle position by moving a grounded metal rod in close proximity to the single, levitated particle. However, this method would sometimes lead to unwanted particles being deposited in the experiment. By using a small, rare-Earth magnet outside of the vacuum chamber, we found nearly identical results without introducing unwanted particles. The magnet was removed in a fraction of a second, while the particle relaxation process took more than 4 s.} 

Two examples of particle trajectories during relaxation to equilibrium after a perturbation, and the corresponding best fit, are shown in Fig.\ \ref{p2}a-b. Assuming $k_c=0$ and ignoring the stochastic noise term, Eqs.\ \ref{eq1}-\ref{eq3} can be solved analytically:
\begin{align}
    w(t) &= A_we^{-\gamma t/2}\cos\left(t\sqrt{\omega_-^2-\frac{\gamma^2}{4}}  + \phi_w\right)
    \label{perturbfit1}\\
    s(t) &= A_se^{-\gamma t/2}\cos\left(t\sqrt{\omega_+^2-\frac{\gamma^2}{4}} + \phi_s\right)
    \label{perturbfit2}\\
    x(t) &= w(t)\cos\theta-s(t)\sin\theta
    \label{perturbfit3}\\
    y(t) &= s(t)\cos\theta + w(t) \sin\theta
    \label{perturbfit4}
\end{align}
Here the fitting parameters $A_w$, $A_s$, $\phi_w$, and $\phi_s$ depend on the initial conditions, and $\gamma$, $\omega_-$, $\omega_+$, and $\theta$ are an estimation of the model parameters as described in Sec.\ \ref{1p_sA}, assuming $k_c$ = 0. 
    
\subsection{Predicting Experimental Data - Results}

We directly compared these measurements with the results from the ML model (the mean of the predictions from neural network and gradient boosting), which measures the parameters \textit{in situ} without perturbations. 
The difference between the model's predictions and the labels inferred from the aforementioned perturbation method lay within the error bars estimated from the simulated test data in parameters $\gamma$, $\omega$, $\delta$ and $\theta$. In other words, the model predicts experimental data as accurately as simulated data, so the mismatch between experimental and simulated data was alleviated. In general, ULI was able to predict $\omega$, $\delta$, and $\theta$, yet with an accuracy that was poor compared to ML, which may be expected since ULI does not require training from multiple datasets. 

Both the perturbation method and ML show excellent agreement with the prediction of $\gamma$ from Epstein's Law (Eq.\ \ref{epstein}). {\color {black}The confinement asymmetry, $\delta$, could be as large as 0.2 although the experimental configuration was quite symmetric and the illuminating laser only contributed to a 1\% asymmetry since a gradient in laser intensity is needed to change the confinement strength. Additionally, the gas flow and pumping rate were low and did not affect $\delta$. We speculate that the asymmetry in the confinement may be due to background flows in the plasma environment. ULI produced wildly varying predictions of $\gamma$, even sometimes negative values. Thus, we did not include it in Fig.\ \ref{p2}c}.

In analyzing the stochastic noise, we found that the prediction of the L\'{e}vy parameter $\alpha$ reflected the particle's effective temperature, $2k_BT\approx m\omega^2(\left<x\right>^2+\left<y\right>^2)$, where $k_B$ is Boltzmann's constant (Fig.\ \ref{p2}g, inset). We have assumed an equipartition between kinetic and potential energy, and expressed the temperature here in terms of the average potential energy to avoid calculating derivatives for the velocity. Importantly, no information about the temperature was passed to the ML model since all time series were normalized. Reported values of $T$ in dusty plasmas driven by Brownian motion vary from 300-1000 K \cite{kong_qiao_matthews_hyde_2016,Himpel2019}. For most experiments, we found $T=300-460$ K, with $1.9<\alpha<2$, indicating nearly Gaussian noise from the room-temperature neutral collisions (Fig.\ \ref{p2}g, inset). Larger temperatures typically corresponded to smaller values of $\alpha$. We speculate that this could be caused by contamination with undetectable, small dust particles since the effective temperature was seen to increase over time in some experiments. Often these particles were ``dropped" by shutting off the plasma, and a new particle was deposited in its place. In any case, the source of the higher effective temperatures was non-Gaussian, although we could not definitively identify the origin of the noise. 

The non-conservative force from Eq.\ \ref{eq3} was smaller than the prediction error bars for most experiments (Fig.\ \ref{p2}h). Part of the motivation for including $k_c$ in our linearized model were observations that particles can undergo small elliptical orbits without any apparent input of energy \cite{Nosenko2020}. In Nosenko et al. \cite{Nosenko2020}, the gravitational leveling of the electrode played a role, presumably due to a feedback between the plate geometry and the background ion flows. Another possibility would be a non-spherical or broken particle, which could then interact with background ion flows.

\section{Two Particle Motion}
\label{2p_s}
\subsection{The Linearized Model}
\begin{figure}
\centering
\includegraphics[width=2.0in]{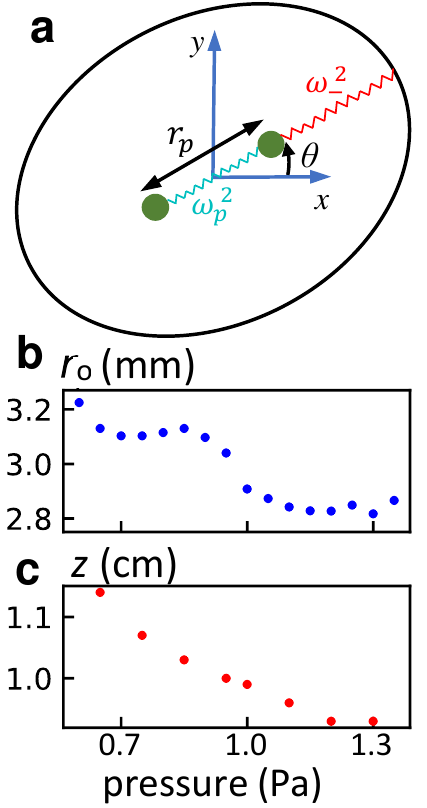}
\caption{(a) Two charged particles experience mutual repulsion, characterized by frequency $\omega_p$ at equilibrium, and external harmonic confinement. $k_c$ and $\omega_+$ are omitted for clarity. All susequent panels show parameter estimates as a function of pressure, $P$. (b-c) The equilibrium particle separation and vertical height above the electrode varied with the gas pressure.}
\label{fig2p}
\end{figure}

When two particles are present, their mutual repulsion, $m\vec{\mathbf{f}}_{ij}=mf_p(r) \vec{\mathbf e}_{ij}$, displaces them from the center of the confining potential, as shown in Fig.\ \ref{fig2p}a. Here, $f_p$ is the reduced force, $\vec{\mathbf e}_{ij}$ is a unit vector from particle $i$ to $j$, $r$ is the particle separation, and $m$ is the mass of a particle. At equilibrium, the particle separation is $r_0$, which varies with pressure and the vertical position $z$ (Fig.\ \ref{fig2p} b-c). In particular, below P $\approx$ 1 Pa, $r_0$ sharply increased to a plateau and the height increased, presumably due to an increase in the plasma Debye length. Here we aim to simultaneous infer:
\begin{align}
    \omega_-^2\frac{r_0}{2}&=f_p(r_0),\label{cond1}\\
    \omega_p^2 &= \left.-\frac{df_p}{dr}\right\rvert_{r=r_0}\label{cond2},
\end{align}
and all other model parameters from Eqs.\ \ref{eq1}-\ref{eq3} with high accuracy using noisy data. We linearize the small-amplitude motion of each particle about their equilibrium position. The equation of motion for particle $i$ is:
\begin{equation}
   \Ddot{\vec{\mathbf r}}_i  = -\vec{\mathbf \nabla}\phi + \vec{\mathbf \nabla} \times \vec{\mathbf A} -  \vec{\mathbf f}_{ij} -\gamma\dot{\vec{\mathbf r}}_i + \vec{\mathbf N}(\alpha),
   \label{eq4}
\end{equation}
\begin{equation}
     \vec{\mathbf f}_{ij} = f_p\vec{\mathbf e}_{ij}=\left(-\omega_p^2 (r - r_0) + \omega^2(1-\delta)\frac{r_0}{2}\right) \vec{\mathbf e}_{ij}.
     \label{eq5}
\end{equation}

\begin{figure}
\centering
\includegraphics[width =3.0in]{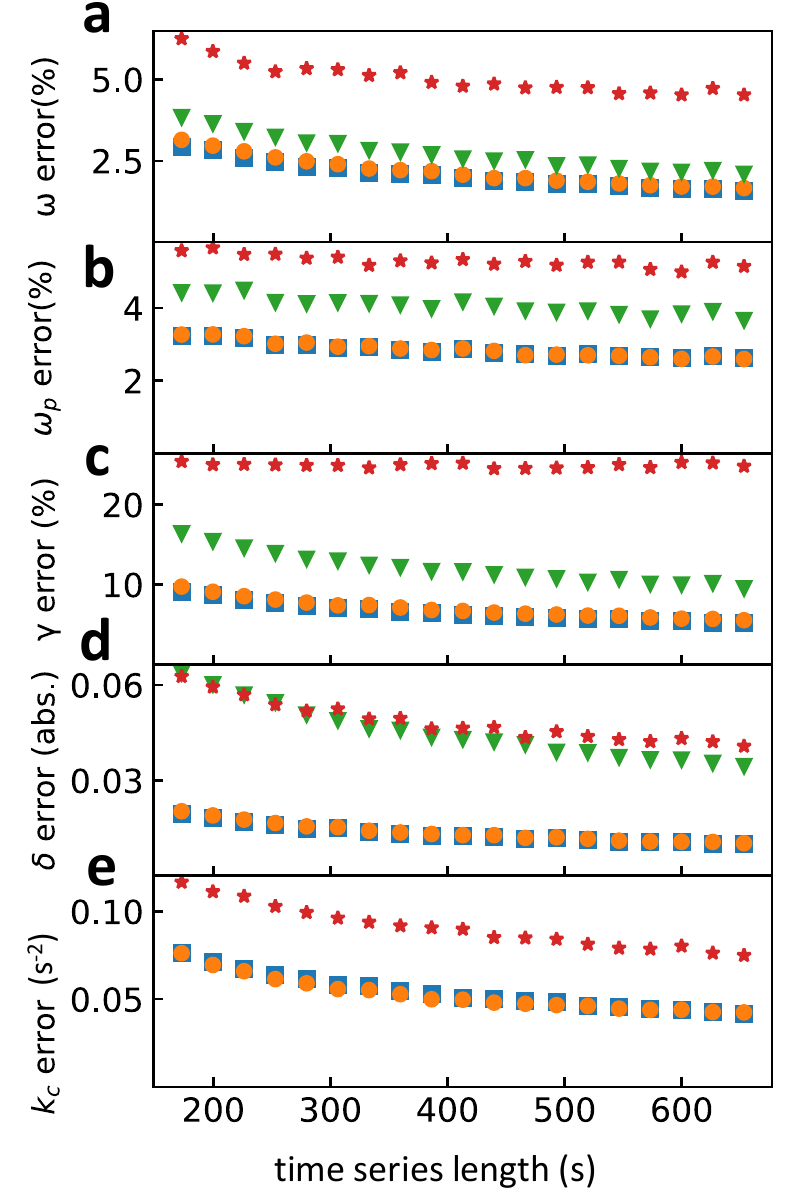}
\caption{The prediction error of various models for multiple parameters on two particle simulated test data. Red stars represent ULI, green triangles represent Fourier spectrum, blue squares represent neural network, and orange circles represent gradient boosting.}
\label{perf_2p}
\end{figure}

The equilibrium force and its differential can be used to solve for 2 independent parameters in a model for $f_p$. The most commonly used model for $f_p$ assumes a screened, Coulomb interaction: 
\begin{equation}
F_D = mf_D=\frac{Q^2}{4\pi\epsilon_0 r }\left(\frac{1}{r} + \frac{1}{\lambda}\right) e^{-r/\lambda}.
\label{debye}
\end{equation}
Here $\epsilon_0$ is the permittivity of free space. With this assumption, similar linearized models have been used to directly infer $Q$ and $\lambda$ from the one-dimensional motion of two particles using Fourier analysis \cite{qiao2017determination,kong2014interaction,mukhopadhyay2012two,Sheridan2007}. Here we allow for entanglement between motion in two dimensions, and provide estimates of all model parameters. Analogous to the single particle model, parameters $\gamma$, $\omega$, $\delta$, $\theta$, $k_c$, $\alpha$, and $\omega_p$ were randomly chosen to simulate time series using Eqs.\ \ref{eq4}-\ref{eq5}. The same feature extraction methods as described for a single particle were applied to the center-of-mass and relative coordinates,  ($\vec{\mathbf{r}}_1+\vec{\mathbf{r}}_2$)/2 and $\vec{\mathbf{r}}_1-\vec{\mathbf{r}}_2$ \cite{supp} with some alterations described in Sec.\ \ref{2p_s}C. The neural network and gradient boosting models were trained on 227 extracted features from 400,000 simulated time series to predict the 7 parameters, and their performance on simulated test data is shown in Fig.\ \ref{perf_2p}. 

\subsection{Simulation Details}

\begin{table*}
\caption{The parameters for 2-particle simulation.}
\label{tab2}
\begin{center}
\begin{tabular}{|c| c| c |c | c|} 
 \hline
 Name & Description & Range & Drift amp. & Drift time \\ 
 \hline\hline
 $\omega$ & confinement freq. & (1.3,2.5) s$^{-1}$ & (0,2\%) & (12,800) s \\ 
 \hline
 $\omega_p$ & interaction freq. & (1.7,3.3) s$^{-1}$ & (0,2\%) & (12,800) s \\ 
 \hline
 $\gamma$ & damping coef.& (0.4,1.7) s$^{-1}$ & (0,2\%) & (12,800) s \\
 \hline
 $\delta$ & asymmetry & (0,0.35) & (0,0.01) & (12,800) s \\
 \hline
 $\theta$ & weak dimension & $(-\pi/2,\pi/2)$ & (0,0.1) & (12,800) s \\
 \hline
 $k_c$ & vortex force coef. & $(-k_{c,max},k_{c,max})$ s$^{-2}$ & None & None \\ 
 \hline
 $\alpha$ & noise distribution & (1.6,2.0) & None & None\\
 \hline
 $c_0$ & noise scale & (0.005,0.03) & None & None\\
 \hline
 $x_0$ & equi. position & 0 & (0,0.02) & (12,800) s\\
 \hline
 $y_0$ & equi. position & 0 & (0,0.02) & (12,800) s\\
 \hline
 $r_0$ & equi. seperation & 1 & (0,0.02) & (12,800) s\\
 \hline
 $m_1$ & mass ratio& (0.9,1.1) & None & None\\
 \hline
 $A$ & 2nd order coef. & (-3,10) s$^{-1}$ & None & None\\
 \hline
 $l_{pixel}$ & simulated pixel width & (0.005,0.02) & None & None\\
 \hline
\end{tabular}
\end{center}
\end{table*}

For simulations involving two particles, the drift of the parameters can change the equilibrium separation of the particles, which is comparable to the amplitude of the Brownian motion. Thus, we allowed for a small drift of many parameters, as listed in Table \ref{tab2}. Importantly, the introduction of a repulsive force between the particles leads to a natural length scale, $r_0$, which is the equilibrium separation between the particles after balancing external confinement and mutual repulsion. In the simulations, $r_0$ is set to 1, but is allowed to drift. This fixed length scale means we must choose the amplitude of the noise to match what is observed in the experiments. The noise scale, $c = c_0 \sqrt{\gamma\omega^2/\Delta t}$, is similar to simulations of for one particle, but here $c_0\ll r_0$. The value of $c_0$ is randomly chosen in each simulation, and represents the amplitude of Brownian motion in units of length measured by the particle separation, $r_0$. Note that $c$ has units of acceleration since mass is normalized and $c_0$ has units of length.

Furthermore, to make the model more general, a second order term with random coefficient $A$ was added to the reduced particle interaction force,
\begin{equation}
f_p= \left(\frac{A}{r_0}(r-r_0)^2-\omega_p^2(r-r_0)+\omega^2(1-\delta)\frac{r_0}{2}\right). 
\end{equation}
This is identical to Eq.\ \ref{eq5}, albeit with the addition of the second order term. Note that typically $r - r_0 \approx c_0 \ll 1$, so the second order term is negligible, but was included for generality. Experimentally, the two particles may be slightly different in size, so a mass difference was considered in simulations. We randomly chose the mass of one particle, $m_1$, to vary by 10\% (Table \ref{tab2}). As a reminder, none of the parameters have a mass unit, so we can arbitrarily fix the sum of the masses, $m_1+m_2=2$. The acceleration of each particle $\Ddot{\vec{\mathbf r}}_i$ was calculated as 
\begin{equation}
    \Ddot{\vec{\mathbf r}}_i = \vec{\mathbf f}_i/m_i
\end{equation}
where $\vec{\mathbf f}_i$ is the sum of all the reduced forces exerted on particle $i$. After each simulation,  pixel-locking noise was added to the trajectory of each particle with $l_{pixel}$, as listed in Table \ref{tab2}. For all simulations, we used a second order, Velocity Verlet time stepping method to integrate the equations of motion.

\begin{figure*}
\centering
\includegraphics[width =\textwidth]{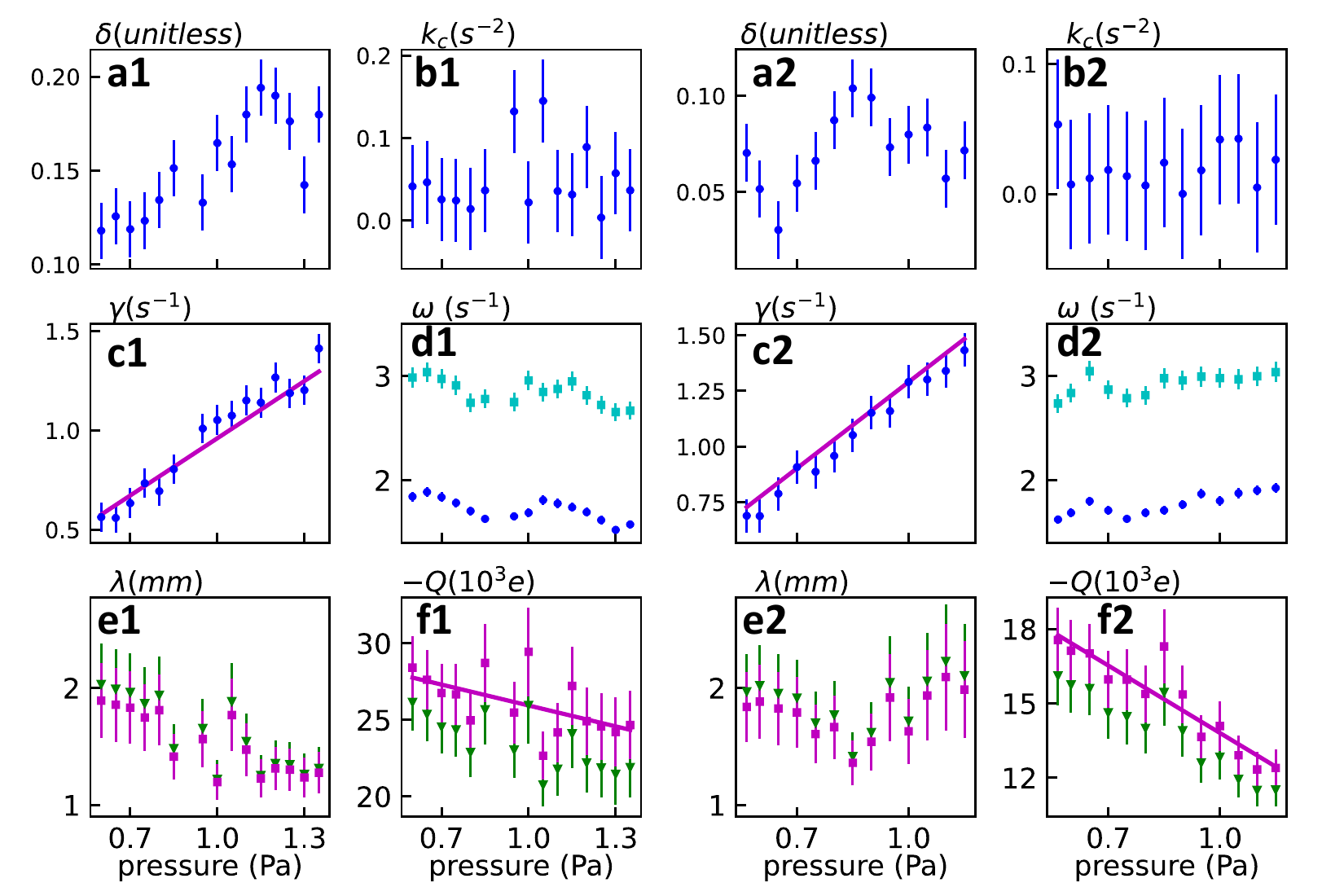}
\caption{ML predictions for the two particle systems. Data is shown for 12.8 $\mu$m particles (denoted by 1) and for 9.46 $\mu$m particles (denoted by 2). The ML prediction is the average of results from the neural network and gradient boosting methods. (a-c) The predictions of $\delta, k_c$, and $\gamma$ by ML are represented by blue circles with error bars estimated from Fig.\ \ref{supp3}. The magenta line in (c) is the theoretical $\gamma$ from Epstein's law. (d) The confinement frequency $\omega$ (blue circles) and interaction frequency $\sqrt{2}\omega_p$ (cyan squares) predicted by ML. (e-f) The Debye length $\lambda$  and particle charge $Q$ calculated from the frequencies using Eqs.\ \ref{debye} (green triangles) and \ref{wake} (magenta squares) from the main text. The magenta line is a linear fit to the magenta squares to guide the eye.}
\label{2p_result}
\end{figure*}

\subsection{Features of two particle motion}
    
    For two particle systems, the rough estimation of $\theta$ is simply the direction in which the particles align (the weak axis), so we don't perform the preprocessing step. The 4D time series representing the motion of the two particles is projected into the center-of-mass motion and the relative motion. The same steps 1-9 as Sec.\ \ref{1p_s}C applies to the center-of-mass motion (126 features), and similar methods are applied to the relative motion with the following revisions: 
    \begin{enumerate}
        \item The fit of the Fourier spectrum only needs 2 projections, parallel and perpendicular to the alignment of the two particles, instead of 4. Furthermore, an additional fit is conducted to the Fourier spectrum of the mode perpendicular to the alignment, where motion predominantly occurs in the azimuthal direction. If the asymmetry $\delta=0$, this mode would have zero frequency. Thus we used a window 0.1 s$^{-1} < \omega < $ 1 s$^{-1}$, and $\gamma_{FT}$, $\omega_{FT}$, but not $A_{FT}$, are recorded as features. Altogether there are 8 features instead of 12.
        \item The same logic applies for autocorrelation, so there are 6 features instead of 8. 
        \item Percentiles and equipartition analysis are performed to 2 (parallel and perpendicular) projections rather than 6. Altogether 4 features instead of 12. 
        
        \item Velocity distribution analysis is performed to 2 projections rather than 3. Altogether 6 features instead of 9
    \end{enumerate}
    The revised 8 analyses gives 46 features. After the aforementioned smoothing, the analyses are performed again for another 46 features. Moreover, since the trajectory is normalized by the particle separation so that $r_0 = 1$, the STD of each of the 4 modes before and after smoothing are 8 new features. Plus the time series length, altogether we have $63\times 2 + 46\times 2 + 8 + 1 = 227$ features for the motion of two particles.


\subsection{Predicting Experimental Data - Results}

Fig.\ \ref{2p_result} shows the ML model prediction on experimental data, for particles with diameters 12.8 $\mu$m (label 1) and 9.46 $\mu$m (label 2). The prediction of $\delta$ is significantly different between the two sizes of particles, as shown in Fig.\ \ref{2p_result}a. This is likely due to their difference in vertical equilibrium position in $z$ (Fig\ \ref{p1_exp}g). Though $k_c$ seems systemically positive for both particles, its amplitude was usually smaller than the error bar (Fig.\ \ref{2p_result}b). We did not specifically train the model on the parameter $\theta$ since it is essentially determined by the alignment of the vector $\vec{\bf r_2}-\vec{\bf r_1}$. Although a finite value of $k_c$ would lead to an angular displacement away from the weak confinement axis, we found that a numerical value of $k_c$ = 0.1 s$^{-2}$ would lead to an small angular displacement of $\approx0.1$ radians.

The damping $\gamma$ varied linearly with $P$ and was well-described by Epstein's law (Eq.\ \ref{epstein}) for both particles, as illustrated in Fig.\ \ref{2p_result}c. The error bars also decreased considerably when compared to the analysis of a single particle (Fig.\ \ref{supp3} vs. Fig.\ \ref{perf_2p}). Both $\omega$ and $\omega_p$ displayed a slight, non-monotonic variation with pressure, outside the range of the error bars in the prediction (Fig.\ \ref{2p_result}d). We note that the eigenfrequency for the relative coordinate motion is $\sqrt{\omega_p^2/\mu + \omega_-^2}$, where $\mu=m_1 m_2/(m_1+m_2)$ is the reduced mass. If the masses of the particles are identical, then $\mu=1/2$. However, assuming that the particles' mass differ by less than 10\%, $\mu = 0.495\pm0.005$. Therefore, the coefficient $\sqrt{1/\mu}\approx\sqrt{2}$ is included in Fig.\ \ref{2p_result}d for clarity.

For 12.8 $\mu$m diameter particles, the ``kink" for $\omega$ and $\omega_p$ lied at the steepest decrease of $r_0$ in Fig.\ \ref{fig2p}. Although these features may be related, we cannot say for sure since the particle height, charge, and Debye length all vary with pressure, and all contribute to $\omega$ and $\omega_p$. Additionally, we found that the parameter $\alpha$ could not be well-predicted from the center-of-mass and relative coordinates in the two particle system. Although we are uncertain why the prediction failed for $\alpha$, we speculate that the distribution in errors from pixel locking are non-additive (unlike the L\'{e}vy stable distribution). The motion of the center-of-mass, for example, is the sum of the particle positions. Thus the total noise distribution for the center-of-mass and relative coordinates could be quite different from one-particle system.


Finally, to estimate $Q$ and $\lambda$, we used two models for $f_p$. The first is Eq.\ \ref{debye}, and the second model is derived from a potential ($mf_w=-d\phi_w/dr$) that incorporates the virtual charge ($\tilde{q}$) from the ion wake beneath each particle \cite{Ishihara1997,kryuchkov2020strange}:
\begin{equation}
    \phi_w = \frac{Q^2}{4\pi\epsilon_0\lambda}\left[\frac{e^{-r/\lambda}}{r/\lambda} - \tilde{q}\dfrac{\lambda e^{-r_w/\lambda}}{r_w+b\lambda e^{-r_w/\lambda}}\right].
    \label{wake}
\end{equation}
Here $r_w = \sqrt{r_0^2+0.12\lambda^2}$, $\tilde{q} = 0.3$, and $b$ = 1. The parameter $b$ controls the size of the virtual charge cloud. After inferring $\omega_p$ and $\omega_-$, the interaction force and it's derivative 
at $r=r_0$ were numerically solved for $Q$ and $\lambda$, following Eqs.\ \ref{cond1} and \ref{cond2}. For both models of the interaction force, the predictions of $Q$ and $\lambda$ differed by only 10\%, as shown in Fig.\ \ref{2p_result}e-f, for both sizes of particles. We expect this difference to be much larger if the particles have significant vertical separation, where ion wake interactions lead to non-reciprocal forces \cite{ivlev2015statistical}. 

For both particles, $Q$ decreased with pressure. The Debye screening length $\lambda$ also decreased with pressure for 12.8 $\mu$m particles, but there seemed to be a kink in $\lambda$ for the 9.46 $\mu$m particle data at $P\approx0.75$ Pa. This is likely due to the variation in height with pressure, and a similar feature is also visible in $\delta$ (Fig.\ \ref{2p_result}a2). It is also important to note that the error in estimating $Q$ and $\lambda$ can be up to 10$\times$ larger than $\omega$ (Fig.\ \ref{2p_result}e-f). This can be illustrated by examining Eqs.\ \ref{cond1}, \ref{cond2}, and \ref{debye}. These equations can be solved analytically, resulting in the following expression for $\lambda$:
\begin{equation}
\lambda=\dfrac{r_0}{2}\left(\dfrac{\omega_p^2+\omega_-^2}{\sqrt{\omega_p^4-\omega_-^4}}-1\right).
\end{equation}
When $\omega_p$ is close to $\omega_-$, perhaps with overlapping error bars, then the uncertainty in $\lambda$ diverges.
This emphasizes the importance of accurate estimation of all parameters, independent of the model chosen for $f_p$. Although we allowed for a 10\% variation in mass between the particles, this variation was not considered when calculating $Q$. The reported values of $Q$ should be taken as an average of the charge on both particles since the mass of each particle may be slightly different.  

Finally, we can compare the results shown in Fig.\ \ref{2p_result}f to theoretical predictions in our plasma conditions. First, we can estimate the total Debye length in the plasma,
\begin{equation}
    \lambda_D=\sqrt{\dfrac{k_B \epsilon_0 T_eT_i}{e^2n_p(T_e+T_i)}},
    \label{debyel}
\end{equation}
where $n_p$ is the quasi-neutral plasma density, $T_i$ is the ion temperature, and $e$ is the elementary charge. As stated previously, $T_e\approx$ 1.3-1.5 eV, and $T_i\approx$ 0.026 eV, so that $\lambda_D$ is dominated by the ion temperature. For our experimental conditions, the plasma density has been measured previously using a Langmuir probe, and $n_p\approx$ $2-5\times10^{13}$ m$^{-3}$ \cite{Harper2020}. Thus, we expect $\lambda_D\approx$ 160-260 $\mu$m. However, the screening length between particles ($\lambda$) is known to be 5-15 times larger than $\lambda_D$ \cite{Nikolaev2016}. Thus, our measurement of 1-2 mm shown in Fig.\ \ref{2p_result}e is reasonable and approximately 10$\times$ larger than $\lambda_D$. Although $\lambda_D$ is expected to increase at low pressures since $n_p$ decreases, the exact dependence of the screening length $\lambda$ on the ion Debye length $\lambda_D$ is not well understood, especially within a plasma sheath. Subsequently, we do not currently have an explanation for the non-monotonic behavior seen in Fig.\ \ref{2p_result}e2.

Second, the particle charge $Q$ can be estimated, to lowest order, by orbital-motion-limited (OML) theory \cite{mott1926theory,Tang2014,Melzer2019b}. In its simplest form, we assume that the electron and ion concentrations are equal (quasi-neutrality). The electron velocity distribution is Maxwellian, but the ion drift velocity $u_i$ towards the electrode must be considered in the ion velocity distribution. In this regime, the dust charge is determined by solving the following equation numerically (see section 2.4.3 in \cite{Melzer2019b}):
\begin{align}
    &0=\sqrt{\dfrac{T_e m_i}{T_im_e }}e^{e Q/4\pi\epsilon_0 a k_B T_e}-\\
    &\frac{\text{erf}(\xi ) \left(2 \pi  \epsilon a k_B T_i  \left(2 \xi ^2+1\right)  -q Q\right)}{8\sqrt{\pi }\epsilon a k_B   T_i \xi }-\frac{1}{2} e^{-\xi ^2},\nonumber
    \label{oml}
\end{align}
where erf($\xi$) is the error function. The parameter $\xi$ is the ratio of the drift velocity to the ion thermal velocity, $\xi=u_i/v_{i,th}$, where $v_{i,th}=\sqrt{2 k_B T_i/m_i}$, $m_{i}$ is the mass of an argon ion, and $m_{e}$ is the electron mass. For a simple estimate, we use the Bohm velocity for the ions because the particles levitate near the edge of the plasma sheath, so that $u_i=\sqrt{ k_B T_e/m_i}$. 

Using the parameters from our experiment, $\xi\approx 5$, and OML theory predicts that the 9.46 $\mu$m particles are negatively charged with 16,600-19,200 electrons, and the 12.8 $\mu$m particles are negatively charged with 22,500-26,000 electrons. This is in excellent agreement with the results shown in Fig.\ \ref{2p_result}f. Additionally, the dust charge should decrease as pressure increases since ion-neutral collisions become more important as the mean free path decreases (\cite{Saitou2018}, \cite{Khrapak2008}, section 2.4.4 in \cite{Melzer2019b}). This expectation is also consistent with Fig.\ \ref{2p_result}f. A more quantitative analysis of the pressure dependence of the particle charge is hampered by small but simultaneous variations of electron temperature, drift velocity, and plasma density as the pressure is changed in the experiment. Additionally, alterations to OML theory to include collisions typically ignore ion drift, and both are present in our experiments. Overall, these results highlight the importance of measuring particle charge \textit{in-situ}, as demonstrated here.

\section{conclusion}
In conclusion, we have overcome two major restrictions when inferring parameters from experimental data with ML: data labeling and mismatch. In our dusty plasma experiments, we label our data by observing particles' relaxation after perturbation. The mismatch between commonly-simulated data (with Gaussian stochastic forces or Gaussian measurement error), and experimental data (with drift, pixel-locking, non-Gaussian stochastic force, etc.) is handled by including these artifacts in our simulation. The agreement between the label for experimental data and our models' prediction demonstrates the alleviation of this mismatch. Additionally, this work provides simultaneous and accurate estimates of multiple parameters from noisy data, for example, the model provides an accurate in-situ determination of Epstein drag, the particle charge, and the Debye length in dusty plasmas. 

Our ML model employed supervised machine learning since stochastic forces and trajectories are an inevitable part of the dynamics. Ideally, the forces in dusty plasmas could be inferred from the experimental data itself, without the need for simulations in the training data set. In dynamical systems with many particles, the interaction force cannot be linearized, yet particle motion is domninated by deterministic forces, and noise is secondary. Thus it may be possible to infer interaction forces in more complex and dynamic systems, which remains an open challenge.

\begin{acknowledgments} 
We thank Ilya Nemenman, Eslam Abdelaleem, and Guram Gogia for stimulating discussions. This material is based upon work supported by the National Science Foundation under Grant No. 2010524, and the U.S. Department of Energy, Office of Science, Office of Fusion Energy Sciences program under Award Number DE-SC0021290.
\end{acknowledgments} 

\bibliography{machine_learning_dusty_plasma}

\end{document}